\documentclass{aa}
\usepackage{rotating}
\usepackage{graphicx}
\usepackage{times}
\begin{document}
\title{An {\it XMM-Newton} observation of the massive binary HD\,159176\thanks{Based on observations with {\it XMM-Newton}, an ESA Science Mission with instruments and contributions directly funded by ESA Member States and the USA (NASA).}}
\author{M.\,De Becker\inst{1} \and G.\,Rauw\inst{1}\thanks{Research Associate FNRS (Belgium)}  \and J.M.\,Pittard\inst{2} \and I.I.\,Antokhin\inst{3}$^,$\inst{4} \and I.R.\,Stevens\inst{5} \and E.\,Gosset\inst{1}$^{\star\star}$ \and S.P.\,Owocki\inst{6}}

\institute{Institut d'Astrophysique, Universit\'e de Li\`ege, All\'ee du 6 Ao\^ut, B\^at B5c, B-4000 Li\`ege (Sart Tilman), Belgium  \and 
Department of Physics \& Astronomy, University of Leeds, Leeds LS2 9JT, UK \and
Department of Physics \& Astronomy, University of Glasgow, Kelvin Building, Glasgow G12 8QQ, Scotland, UK \and 
On leave from: Sternberg Astronomical Institute, Moscow University, Universitetskij Prospect, 13, Moscow 119899, Russia \and
School of Physics \& Astronomy, University of Birmingham, Edgbaston 
Birmingham B15 2TT, UK \and
Bartol Research Institute, University of Delaware, Newark, DE 19716, USA}

\date{Received date / Accepted date}
\authorrunning{M.\,De Becker et al. }
\titlerunning{An {\it XMM-Newton} observation of HD\,159176}

\abstract{We report the analysis of an {\it XMM-Newton} observation of the close binary HD\,159176 (O7\,V + O7\,V). The observed L$_\mathrm{X}$/L$_\mathrm{bol}$ ratio reveals an X-ray luminosity exceeding by a factor $\sim$7 the expected value for X-ray emission from single O-stars, therefore suggesting a wind-wind interaction scenario. EPIC and RGS spectra are fitted consistently with a two temperature {\tt mekal} optically thin thermal plasma model, with temperatures ranging from $\sim$2 to 6\,10$^6$ K. At first sight, these rather low temperatures are consistent with the expectations for a close binary system where the winds collide well before reaching their terminal velocities. We also investigate the variability of the X-ray light curve of HD\,159176 on various short time scales. No significant variability is found and we conclude that if  hydrodynamical instabilities exist in the wind interaction region of HD\,159176, they are not sufficient to produce an observable signature in the X-ray emission. Hydrodynamic simulations using wind parameters from the literature reveal some puzzling discrepancies. The most striking one concerns the predicted X-ray luminosity which is one or more orders of magnitude larger than the observed one. A significant reduction of the mass loss rate of the components compared to the values quoted in the literature alleviates the discrepancy but is not sufficient to fully account for the observed luminosity. Because hydrodynamical models are best for the adiabatic case whereas the colliding winds in HD\,159176 are most likely highly radiative, a totally new approach has been envisaged, using a geometrical steady-state colliding wind model suitable for the case of radiative winds. This model successfully reproduces the spectral shape of the EPIC spectrum, but further developments are still needed to alleviate the disagreement between theoretical and observed X-ray luminosities.
\keywords{Stars: early-type -- Stars: individual: HD\,159176 -- Stars: winds, outflow -- Stars: binaries: general -- X-rays: stars}}
\maketitle

\section{Introduction}
HD\,159176 is a relatively bright (m$_v = 5.7$) double-lined spectroscopic binary (Trumpler \cite{Tr}) in the young open cluster NGC\,6383. The system has been well studied in the visible and UV wavelengths. Conti et al.\ (\cite{CCJ}) derived an orbital solution that was subsequently improved by Seggewiss \& de Groot (\cite{SdG}), Lloyd Evans (\cite{LE}) and most recently by Stickland et al.\ (\cite{SKPP}). The binary has an orbital period of 3.367\,days and consists of two nearly identical O-stars in a circular orbit.

Conti et al.\ suggested that both stars are O7 stars that have evolved off the main-sequence and nearly fill up their Roche lobes. However, Stickland et al.\ argued that the stars were probably not evolved. Although the system does not display photometric eclipses, Thomas \& Pachoulakis (\cite{TP}) reported ellipsoidal variability with an amplitude of about 0.05\,mag in the optical and UV wavebands. These light curves were analysed by Pachoulakis (\cite{Pach}) who inferred radii of both stars of order $0.25\,a$, where $a$ is the orbital separation. According to these results, the stars do not fill up their critical volume and are thus not very deformed.\\ 

Several observations point towards the existence of a wind interaction process in HD\,159176. For instance, the optical spectrum of the system displays the so-called Struve-Sahade effect, i.e.\ the absorption lines of the approaching star appear stronger (Conti et al.\ \cite{CCJ}, Seggewiss \& de Groot \cite{SdG}, Lloyd Evans \cite{LE}), although the reverse effect is seen in the UV (Stickland et al.\ \cite{SKPP}). Though the origin of this effect is as yet not established, a commonly proposed scenario involves the existence of an interaction process within the binary system (Gies et al.\ \cite{gies}). 

Analysing UV resonance line profiles of HD\,159176 as observed with {\it IUE}, Pachoulakis (\cite{Pach}) derived a mass-loss rate of about $3\,10^{-6}$\,M$_{\odot}$\,yr$^{-1}$ for each star (note that this value is a factor five larger than the one derived by Howarth \& Prinja \cite{HP}). The stellar winds of the components of HD\,159176 are therefore probably sufficiently energetic to interact. The interaction region is expected to be located roughly mid-way between the stars and it prevents the wind of each star from deploying into the direction towards the other star. This situation has an impact on the UV resonance line profiles and the UV light curve (Pachoulakis \cite{Pach}, Pfeiffer et al.\ \cite{PPKS}). Pfeiffer et al.\ analysed the variations of the UV resonance lines and concluded that there exists a source of extra emission between the stars. These authors associated this extra emission with resonant scattering of photospheric light by the material inside a colliding wind region. According to their analysis, this shock region is slightly wrapped around the secondary star. 

HD\,159176 was detected as a rather bright X-ray source with {\it EINSTEIN} ($0.157 \pm 0.008$\,cts\,s$^{-1}$ with the IPC, Chlebowski et al.\ \cite{Chle}) and {\it ROSAT} ($0.291 \pm 0.034$\,cts\,s$^{-1}$ during the All Sky Survey, Bergh\"ofer et al.\ \cite{BSC}). Chlebowski \& Garmany (\cite{CG}) suggested that the excess X-ray emission observed in many O-type binaries compared to the expected intrinsic contribution of the individual components is produced by the collision of the stellar winds (see also e.g.\ Stevens et al.\ \cite{SBP}). Therefore, it seems likely that at least part of the X-ray flux of HD\,159176 may originate in the wind interaction region. In contrast with this picture, Pfeiffer et al.\ (\cite{PPKS}) suggest that the bulk of the X-ray emission arises primarily from the intrinsic emission of the individual components, rather than from a colliding wind interaction.\\

To clarify this situation, we obtained an AO1 {\it XMM-Newton} observation of HD\,159176. Since the system is X-ray bright, it is well suited to investigate the X-ray properties of a short-period early-type binary.

\begin{table}[h]
\caption{\label{tbl-1a} Relevant parameters of the HD\,159176 binary system adopted throughout this paper unless otherwise stated. The numbers are taken from Pachoulakis (\cite{Pach}, P96) and Diplas \& Savage (\cite{DS}, DS).}
\begin{center}
\begin{tabular}{l c c c}
\hline
Parameter & Prim. & Sec. & Ref. \\
\hline
$a\,\sin{i}$\,(R$_{\odot}$) & \multicolumn{2}{c}{28.9} & P96 \\
$M$\,(M$_{\odot}$) & 31.9 & 31.6 & P96 \\
$i\,(^{\circ})$ & \multicolumn{2}{c}{$\sim 50$} & P96 \\
$T_{\rm eff}$\,(K) & 42500 & 35000 & P96 \\
$R_*$\,(R$_{\odot}$) & 9.8 & 9.3 & P96 \\
$v_{\infty}$\,(km\,s$^{-1}$) & \multicolumn{2}{c}{2850} & P96 \\
${\dot{\it {\rm M}}}$\,(M$_{\odot}$\,yr$^{-1}$) & $3.2\,10^{-6}$ & $2.6\,10^{-6}$ & P96 \\
$\log{\it {\rm N_{\rm H, ISM}}}$\,(cm$^{-2}$) & \multicolumn{2}{c}{21.23} & DS\\
\hline
\end{tabular}
\end{center}
\end{table}

\section{Observations \label{sect_obs}}
The observation of HD\,159176 with {\it XMM-Newton} (Jansen et al.\,\cite{xmm}) took place during revolution 229, in March 2001 (JD\,2451977.903 -- 2451978.338). According to the ephemeris given by Stickland et al.\ (\cite{SKPP}), the whole observation covered phases between 0.53 and 0.66 (phase 0.0 corresponding to the maximum radial velocity of the primary component).
 
The total observation was split into three exposures. During the first exposure (Obs.\ ID 0001730401), the satellite was still inside the radiation belts and only the RGS instruments were used for on-target observations. The total on-target EPIC exposure time of the other two exposures (Obs.\ IDs 0001730201 and 0001730301) is about 26\,ks.

The three EPIC instruments were operated in the full frame mode (Turner et al.\,\cite{mos}, Str\"uder et al.\,\cite{pn}). All three EPIC instruments used the thick filter to reject optical light. The two RGS instruments were operated in the default spectroscopy mode (den Herder et al.\,\cite{RGS}).\\

We used version 5.2 of the {\it XMM-Newton} Science Analysis System ({\sc sas}) to reduce the raw EPIC data. For each EPIC camera, the observation data files (ODFs) of the two exposures were merged by the {\it XMM-Newton} Science Operation Center (SOC) into a single set of observation data files. In order to process the EPIC-MOS files with the `emproc' pipeline chain, we had to manually correct some keywords in the ODF summary file generated by the `odfingest' task. For the pn camera, the merged ODF had to be reprocessed by the {\it XMM-Newton} SOC to generate an event list.

The only events considered in our analysis were those with pattern 0 -- 12 for EPIC-MOS and 0 for EPIC-pn respectively (Turner et al.\,\cite{mos}). We found no event with pattern 26 -- 31, so that we conclude that no significant pile-up affects our data.

In this paper, we discuss only the X-ray data for HD\,159176. The other X-ray sources found in the EPIC field of view and associated with the open cluster NGC\,6383 are discussed in a separate paper (Rauw et al.\,\cite{paper2}). The response matrices used for the EPIC spectra of HD\,159176 were those provided by the SOC. No difference was found between spectra obtained with these response matrices and those obtained with the {\sc sas}-generated ones.

For the RGS data, separate ODFs were provided for each of the three observations. The three data sets yield exposure times of respectively about 10, 16 and 9\,ks. The RGS data were processed with version 5.3 of the {\sc sas}. The raw data were run through the `rgsproc' pipeline meta-task. Appropriate response matrices were generated using the `rgsrmfgen' {\sc sas} task.

\section{Dealing with high background level episodes}

We extracted a light curve at very high energies (Pulse Invariant (PI) channel numbers $>$ 10\,000) that revealed a high level of soft proton background. The soft protons responsible for these flares are thought to be accelerated by magnetospheric reconnection events unrelated to solar flares (Lumb \cite{lumb}). Unfortunately this high-level background affects a significant fraction of the exposures. Indeed, nearly one half of the total EPIC observation is affected. For this reason, one has to check whether a bad time interval rejection is necessary, or whether the background correction of spectra could take these undesired events properly into account. As a convincing test, we generated EPIC spectra on the basis of the whole event lists (including events occurring during the flare). On the other hand, we filtered the event lists for time intervals with high-energy (PI $>$ 10\,000) count rates exceeding thresholds of 0.20 and 1.10\,cts\,s$^{-1}$, for the MOS and pn cameras respectively, in order to reject the flares and obtain filtered spectra. Of course, the spectra obtained with the complete event list have a better signal-to-noise ratio than the spectra obtained from the filtered one because of the longer exposure time (about 25 ks for the whole exposure, instead of about 13 ks for the filtered data).\\
Following detailed comparison of the results from the two approaches, we did not find any significant difference between the fitting results (see Section\,\ref{sect_epic}) relevant to both cases. For this reason, we decided not to reject the flare intervals from our EPIC data set, and the effective exposure times (see Table\,\ref{expos}) are thus close to the performed observation duration.\\

\begin{table}
\caption{\label{expos}Performed and useful exposure times for all {\it XMM-Newton} instruments. The useful RGS exposure time is considerably reduced due to a strong flare of soft protons. The fourth column yields the background-corrected count rate for the source, between 0.4 and 10 keV for EPIC, and between 0.3 and 2.5 keV for RGS. In the case of the RGS instruments, only exposure 0001730201 is considered. The error bars on the count rate represent the $\pm$ 1\,$\sigma$ standard deviation.}
\begin{center}
\begin{tabular}{cccc}
\hline
Instrument	& Performed & Effective & Background-corrected \\
	& duration & exposure & count rate \\
	& (s) & (s) & (count\,s$^{-1}$) \\
\hline
MOS1 & 25\,901 & 25\,435 & 0.512 $\pm$ 0.005 \\
MOS2 & 25\,906 & 25\,473 & 0.532 $\pm$ 0.005 \\
pn & 22\,759 & 20\,402 & 1.303 $\pm$ 0.009 \\
RGS1 & 16\,495 & 7490 & 0.085 $\pm$ 0.006 \\
RGS2 & 16\,535 & 6838 & 0.101 $\pm$ 0.006 \\
\hline
\end{tabular}
\end{center}
\end{table}

\begin{figure}
\begin{center}
\resizebox{8.5cm}{5.5cm}{\includegraphics{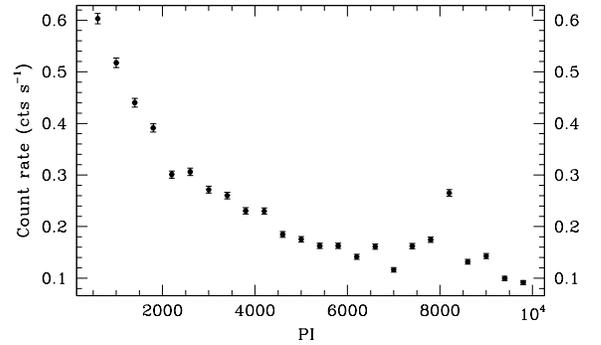}}
\caption{EPIC-pn count rates versus PI for the global background (including the soft proton flares) for a $\sim$ 350 arcmin$^2$ area (about a half of the whole detector plane area). The background is much stronger in low energy bands than at higher energies. The bump seen near PI = 8000 is due to fluorescent emission lines produced by charged particles interacting with the camera body material. The error bars represent the $\pm$ 1\,$\sigma$ standard deviation. \label{flare}}
\end{center}
\end{figure}

Timing analysis of the flare events was also performed to evaluate their impact on the lower energy (PI$<$10\,000) light curves. We extracted EPIC-pn events with boxes covering the 12 CCDs, but excluding the sources appearing in the field. Light curves were extracted, between PI 400 and 10\,000, with a PI width of 400 units and time bins ranging from 50 to 1000 s. These light curves show that the mean level of flare contribution to the background is only about twice the mean level of the proton unaffected background. Fig.\,\ref{flare} shows the count rate versus PI, where each point stands for a 400 PI channel interval, for the time interval affected by the main flare. The count rate of this background decreases by about a factor 10 from PI 400 to 10\,000. The bump near PI 8000 is due to fluorescent emission lines produced by interaction of charged particles with the camera body material (Lumb \cite{lumb}). Comparing the overall background level (including the flare) to the count rate in the source region over PI $\in$ [400:800], i.e. over the energy range with the highest background level, we find that, at its peak level, the total background contributes only for about 0.9\,\% of the corrected source count rate. The same calculations performed in other energy bands reveal that the background contribution below 5 keV is always at least 4 times lower than the source count rate. This result confirms that the rejection of the soft proton time intervals is irrelevant for the analysis of HD\,159176, and that the flare is properly taken into account by the background subtraction step.\\

\begin{figure}
\begin{center}
\resizebox{8.5cm}{5.5cm}{\includegraphics{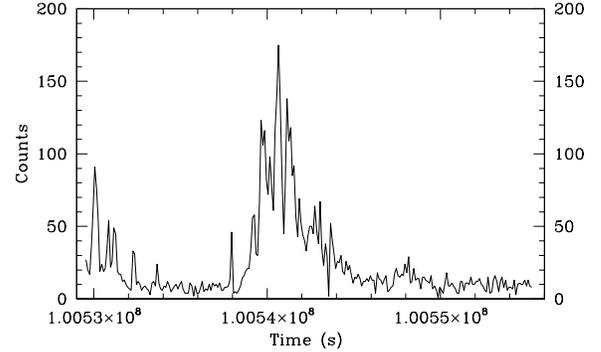}}
\caption{Light curve for events with PI $>$ 10\,000 obtained for the EPIC-MOS2 field and covering the two exposures where the EPIC instruments were operating. A time bin of 100\,s was adopted. \label{lcflare}}
\end{center}
\end{figure}

However, this conclusion does not hold for RGS data. Indeed, the comparison of spectra obtained with and without the flare epochs revealed significant differences. As a consequence, low-background good time intervals were selected by application of count thresholds to the high energy light curves. These thresholds were evaluated, on the basis of a visual inspection of the light curves. The third (9\,ks) exposure was almost completely contaminated by high background, whereas the first (10\,ks) observation gave only about 3\,ks of useful time. Only the 16\,ks exposure could be used, although more than half of it had also to be rejected. For this reason, the effective RGS exposure time is severely reduced, as shown in Table\,\ref{expos}.\\

\section{The EPIC spectrum of HD\,159176\label{sect_epic}}
The source spectrum was extracted within a circular region of 60\,arcsec radius. The intersection with a circular region of radius 15\,arcsec centered on a faint point source located at RA=17:34:40 and DEC=--32:34:33 (Equinox J2000) was rejected. The background was selected within an annular region centered on HD\,159176, with an outer radius 1.414 times larger than for the source region. Again, the intersections with 15\,arcsec circles centered on two point sources (at RA=17:34:48, DEC=--32:34:24 and RA=17:34:48, DEC=--32:35:21) were rejected. Consequently, source and background regions extend nearly over the same area. We also tried to select alternative background regions with boxes distributed near the source region, and avoiding the numerous point sources of the field. This selection did not give a significant difference in the background corrected spectrum. In the case of the pn camera, an additional difficulty arises from the gaps between the CCDs of the pn detector. We excluded the gaps manually from the source and background regions.\\
The additional figure (jpeg format) shows the EPIC-MOS1 field with the selected regions for the source and the background in the annular case. The small circular regions are used for the exclusion of the point sources located inside these source and background regions.\\

\begin{figure}
\begin{center}
\resizebox{8.5cm}{5.0cm}{\includegraphics{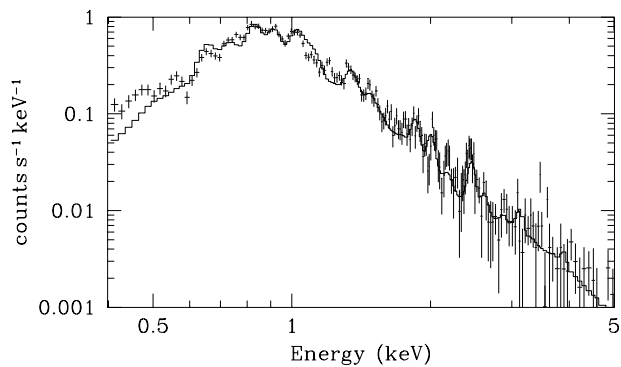}}
\caption{EPIC-MOS1 spectrum of HD\,159176 fitted with the two-temperature {\tt mekal} model between 0.4 and 5.0 keV.\label{m1spec}}
\end{center}
\end{figure}

The emission lines  of the RGS spectrum (see the next section) indicate that at least the low energy X-ray emission of HD\,159176 is dominated by a thermal plasma emission. On the other hand, the shock region in a colliding wind binary might accelerate free electrons up to relativistic velocities and in combination with the intense UV radiation fields of the two stars, these particles could produce a hard non-thermal X-ray emission through inverse Compton scattering (see Chen \& White\,\cite{CW}, Eichler \& Usov\,\cite{EU}). Therefore, we tested various models combining optically thin thermal plasma models with a non-thermal power law component. In the framework of a model where the X-ray emission arises from shocks (either in the wind interaction or in instabilities of the individual winds), one should observe a range of plasma temperatures corresponding to the distribution of the pre-shock wind velocities. Therefore, one expects a priori that more than one thermal component is required to fit the observed spectrum. So, the spectra have been fitted either with two temperature {\tt mekal} thermal plasma models (Mewe et al.\,\cite{mewe}; Kaastra \cite{ka}) or with models containing a single temperature {\tt mekal} component along with a non-thermal power law component. For each thermal component, solar metallicity was assumed.\\
The spectra were rebinned to achieve a minimum of 9 and 25 counts per energy bin for the MOS and pn data respectively and then fitted with the {\sc xspec} software (version 11.1.0). For each model, we used an interstellar neutral hydrogen column density of N$_\mathrm{H}$ = 0.17\,10$^{22}$\,cm$^{-2}$ (Diplas \& Savage \cite{DS}), and a second absorption component to account for circumstellar absorption. The results are listed in Table\,\ref{fits}. We note that the best-fitting power law components turn out to be rather steep with photon indices ($\Gamma$) above 3.8. This situation is very much reminiscent of the results obtained for the O4\,V star 9\,Sgr (see Rauw et al.\ \cite{9sgr}). Figure \ref{m1spec} illustrates the spectrum of HD\,159176 as observed with the MOS1 instrument. Note that including different absorption columns for the individual model components does not improve the quality of the fits. In fact, {\sc xspec} fails to fit at least one of the column densities.\\

\begin{table*}
\caption{Fitted model parameters for EPIC spectra of HD\,159176. The upper part of the table concerns the {\tt wabs$_\mathrm{ISM}$*wabs$_\mathrm{WIND}$*(mekal + mekal)} model, whilst the bottom part summarizes the results for the {\tt wabs$_\mathrm{ISM}$*wabs$_\mathrm{WIND}$*(mekal + power)} model. The N$_\mathrm{H}$ given in the table refers to the circumstellar (i.e.\ wind) column assumed to be neutral. The interstellar column density is frozen at 0.17\,10$^{22}$\,cm$^{-2}$ for all models. The Norm parameter of the {\tt mekal} component is defined as $(10^{-14}/(4\,\pi\,D^2))\int{n_\mathrm{e}\,n_\mathrm{H}\,dV}$, where $D$, $n_\mathrm{e}$ and $n_\mathrm{H}$ are respectively the distance to the source (in cm), and the electron and hydrogen densities (in cm$^{-3}$), whereas for the power law this parameter corresponds to the photon flux at 1 keV. The error bars represent the 1 $\sigma$ confidence interval.\label{fits}}
\begin{center}
\begin{tabular}{ccccccc}
\hline
\hline
	& N$_\mathrm{H}$ & kT$^1$ & Norm$^1$ & kT$^2$ & Norm$^2$ & $\chi^2_\nu$ \\
	& (10$^{22}$ cm$^{-2}$) & (keV) & & (keV) & & d.o.f. \\
\hline
\hline
MOS1 & 0.41 & 0.22 & 2.04 10$^{-2}$ & 1.05 & 1.51 10$^{-3}$ & 2.19 \\
	& $\pm$ 0.03 & $\pm$ 0.04 & $\pm$ 0.38 10$^{-2}$ & $\pm$ 0.02 & $\pm$ 0.09 10$^{-3}$ & 197 \\
\hline
MOS2 & 0.39 & 0.21 & 2.11 10$^{-2}$ & 1.00 & 1.58 10$^{-3}$ & 2.65 \\
	& $\pm$ 0.03 & $\pm$ 0.04 & $\pm$ 0.40 10$^{-2}$ & $\pm$ 0.02 & $\pm$ 0.09 10$^{-3}$ & 198 \\
\hline
pn & 0.35 & 0.18 & 1.88 10$^{-2}$ & 0.59 & 2.98 10$^{-3}$ & 1.87 \\
	& $\pm$ 0.02 & $\pm$ 0.03 & $\pm$ 0.26 10$^{-2}$ & $\pm$ 0.01 & $\pm$ 0.17 10$^{-3}$ & 321 \\
\hline
\hline
	& N$_\mathrm{H}$ & kT & Norm$^1$ & $\Gamma$ & Norm$^2$ & $\chi^2_\nu$ \\
	& (10$^{22}$ cm$^{-2}$) & (keV) & & & & d.o.f. \\
\hline
\hline

MOS1 & 0.26 & 0.30 & 4.69 10$^{-3}$ & 4.01 & 1.87 10$^{-3}$ & 1.65 \\
	& $\pm$ 0.02 & $\pm$ 0.01 & $\pm$ 0.84 10$^{-3}$ & $\pm$ 0.09 & $\pm$ 0.18 10$^{-3}$ & 197 \\
\hline
MOS2 & 0.24 & 0.29 & 5.08 10$^{-3}$ & 3.80 & 1.69 10$^{-3}$ & 1.99 \\
	& $\pm$ 0.02 & $\pm$ 0.01 & $\pm$ 0.89 10$^{-3}$ & $\pm$ 0.08 & $\pm$ 0.16 10$^{-3}$ & 198 \\
\hline
pn & 0.28 & 0.28 & 6.72 10$^{-3}$ & 4.09 & 1.67 10$^{-3}$ & 1.70 \\
	& $\pm$ 0.01 & $\pm$ 0.01 & $\pm$ 0.80 10$^{-3}$ & $\pm$ 0.08 & $\pm$ 0.14 10$^{-3}$ & 321 \\
\hline
\hline
\end{tabular}
\end{center}
\end{table*}

\begin{figure}
\begin{center}
\resizebox{8.5cm}{5.0cm}{\includegraphics{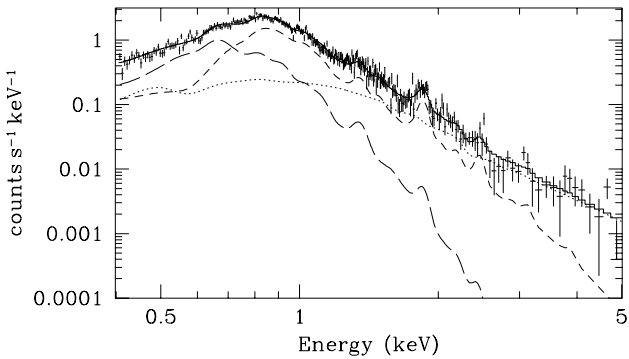}}
\end{center}
\caption{EPIC-pn spectrum of HD\,159176 fitted with the two {\tt mekal} components + power law model between 0.4 and 5.0\,keV. The three components are also individually displayed. The power law begins to dominate the spectrum at energies above 2 keV.\label{newfit}}
\end{figure}

Another model was fitted with some success to the EPIC-pn data. This model consists of three components: 2 thermal {\tt mekal} components and a non-thermal power law component (see Sect.\,\ref{concl} for a possible interpretation of this component). This fit yields a better $\chi_{\nu}^2$, albeit the value obtained for the photon index is once more rather large. The results are summarized in Table\,\ref{fitpn} for the two background regions selected. One can see that both background techniques yield very similar results. This comforts us with the idea that background selection is not a critical issue for this source, and that the annular region with parasite sources excluded provides a satisfactory description of the actual background. Figure\,\ref{newfit} displays the pn spectrum along with the best-fit corresponding to this latter model. We note that this three component model yields a significantly better fit to the EPIC-pn spectrum above 2 keV than the two component models.\\

\begin{table*}
\caption{Fitted model parameters for the EPIC-pn spectrum of HD\,159176 for two different background estimation techniques. The adopted model is {\tt wabs$_\mathrm{ISM}$*wabs$_\mathrm{WIND}$*(mekal + mekal + power)}. The column densities given in the table refer to the circumstellar material. The interstellar column density is frozen at 0.17\,10$^{22}$ cm$^{-2}$. The error bars represent the 1 $\sigma$ confidence interval.\label{fitpn}}
\begin{center}
\begin{tabular}{ccccccccc}
\hline
\hline
	& N$_\mathrm{H}$ & kT$^1$ & Norm$^1$ & kT$^2$ & Norm$^2$ & $\Gamma$ & Norm$^3$ & $\chi^2_\nu$ \\
	& (10$^{22}$ cm$^{-2}$) & (keV) & & (keV) & & & & d.o.f. \\
\hline
\hline
Annulus & 0.20 & 0.20 & 4.39 10$^{-3}$ & 0.58 & 1.79 10$^{-3}$ & 3.50 & 7.35 10$^{-4}$ & 1.11 \\
	& $\pm$ 0.02 & $\pm$ 0.01 & $\pm$ 1.05 10$^{-3}$ & $\pm$ 0.01 & $\pm$ 0.19 10$^{-3}$ & $\pm$ 0.12 & $\pm$ 1.25 10$^{-4}$ & 319 \\
\hline
Boxes & 0.18 & 0.21 & 3.66 10$^{-3}$ & 0.59 & 1.72 10$^{-3}$ & 3.43 & 8.12 10$^{-4}$ & 1.09 \\
	& $\pm$ 0.01 & $\pm$ 0.01 & $\pm$ 0.83 10$^{-3}$ & $\pm$ 0.01 & $\pm$ 0.18 10$^{-3}$ & $\pm$ 0.10 & $\pm$ 1.13 10$^{-4}$ & 319 \\
\hline
\hline
\end{tabular}
\end{center}
\end{table*}

The main motivation behind these fits is to provide an overall description of the observed spectral distribution that can then be used to derive fluxes and to check for consistency with previous observations. With the parameter values obtained for these fits, we can determine the flux emitted by HD\,159176 between 0.4 and 10.0 keV. For this purpose we adopt the best fitting model obtained for EPIC-pn, i.e. the three component model whose parameters are shown in Table\,\ref{fitpn}. The flux corrected for interstellar absorption only is 5.52\,10$^{-12}$ erg\,cm$^{-2}$\,s$^{-1}$. If we adopt a distance to NGC\,6383 of 1.5 kpc, we obtain a luminosity of 1.48\,10$^{33}$ erg\,s$^{-1}$. This value can be compared to X-ray luminosities obtained with {\it EINSTEIN} (Chlebowski et al.\,\cite{Chle}) and {\it ROSAT} (Bergh\"ofer et al.\,\cite{BSC}). In the case of {\it EINSTEIN}, Chlebowski et al.\ converted the IPC count rates between 0.2 and 3.5\,keV into fluxes, assuming a 0.5\,keV thermal bremsstrahlung model and an absorption column (0.209\,10$^{22}$ cm$^{-2}$) calculated on the basis of the color excess E$(B-V)$ determined through a relation given by Mihalas \& Binney (\cite{MB}) for an intrinsic color index $(B-V)_0$ of $-$0.309 and an observed $(B-V)$ color index of $+$0.04. For {\it ROSAT}, Bergh\"ofer et al.\ converted the PSPC count rate (between 0.1 and 2.4\,keV) into a flux assuming a 0.73\,keV Raymond \& Smith (\cite{RS}) optically thin plasma model. The same absorption column as for {\it EINSTEIN} was used. The X-ray luminosities scaled to the 1.5 kpc distance are shown in the second column of Table\,\ref{lum} for {\it EINSTEIN}, {\it ROSAT}, and {\it XMM}.\\
The luminosities derived from the different satellites are quite different. These discrepancies cannot be explained by the differences in the sensitivity ranges of the various instruments. Instead, it seems more likely that the model assumptions used to convert the {\it EINSTEIN}-IPC and {\it ROSAT}-PSPC count rates into luminosities are responsible for the discrepant luminosities. To check this, we have used {\sc xspec} to simulate {\it EINSTEIN}-IPC and {\it ROSAT}-PSPC spectra of HD\,159176. The three component model derived from the fit of the EPIC-pn data (see Table 4) was folded through the response matrices of the IPC and PSPC instruments. The corresponding theoretical count rates, 0.138\,cts\,s$^{-1}$ for the IPC and 0.287\,cts\,s$^{-1}$ for the PSPC, are in very good agreement with the observed values (see Table 5). Therefore, we conclude that the archive data do not reveal evidence for long-term variability of the X-ray flux of HD\,159176.\\
Since the orbit of HD\,159176 does not have a significant eccentricity, it is expected that a possible orbital modulation of the X-ray emission from a colliding wind interaction zone would be due entirely to the changing circumstellar column density along the line of sight. However, given the rather low inclination ($i \sim 50^{\circ}$, Table\,1), we expect only a marginal orbital modulation of the X-ray flux at energies above 0.4\,keV. At this stage, it is worth recalling that our {\it XMM} observation of HD\,159176 was obtained during orbital phases 0.53--0.66, i.e.\ shortly after quadrature. Therefore, we expect no strong occultation effect of the X-ray emission from either component or from the wind interaction region: our value of the luminosity should thus reflect the typical X-ray emission of HD\,159176.\\
\linebreak
We use the total bolometric luminosity of HD\,159176 as derived from the parameters in Table\,\ref{tbl-1a}, i.e.\ 1.49\,10$^{39}$ erg\,s$^{-1}$, to compute the L$_\mathrm{X}$/L$_\mathrm{bol}$ ratios given in Table\,\ref{lum}. On the other hand, we estimate the X-ray luminosity of the individual components by means of the empirical relation of Bergh\"ofer et al.\,(\cite{BSDC}). The ratio of the sum of the expected X-ray luminosities and the total bolometric luminosity amounts to 1.48\,10$^{-7}$, which is about a factor 7 inferior to the {\it XMM} value listed in Table\,\ref{lum}, suggesting that there exists indeed a moderate excess X-ray emission in HD\,159176.\\

\begin{table}[h]
\caption{\label{lum} X-ray luminosities and luminosity ratios from {\it EINSTEIN}, {\it ROSAT} and {\it XMM} observations of HD\,159176. Luminosities are calculated for a 1.5 kpc distance. The Obs. CR column gives the count rates from Chlebowski et al. (\cite{Chle}) and Bergh\"ofer et al. (\cite{BSC}) respectively for {\it EINSTEIN} and {\it ROSAT}, as well as our value for EPIC-pn. The Th. CR column provides the theoretical count rates calculated on the basis of the individual instrument response matrices for the 3 component model of Table\,\ref{fitpn} (see text). }
\begin{center}
\begin{tabular}{c c c c c}
\hline
Satellite	& L$_\mathrm{X}$	& L$_\mathrm{X}$/L$_\mathrm{bol}$ & Obs. CR & Th. CR\\
	& (erg\,s$^{-1}$)	&  & (cts\,s$^{-1}$) & (cts\,s$^{-1}$)\\
\hline
{\it EINSTEIN}	& 4.33\,10$^{33}$ & 2.89\,10$^{-6}$ & 0.157 & 0.138\\
{\it ROSAT}	& 8.43\,10$^{32}$ & 5.64\,10$^{-7}$ & 0.291 & 0.287\\
{\it XMM}	& 1.48\,10$^{33}$ & 9.93\,10$^{-7}$ & 1.303 & 1.328\\
\hline
\end{tabular}
\end{center}
\end{table} 

\begin{table*}[t]
\caption{Model parameters for a two-temperature {\tt mekal} model fitted to the RGS1 and RGS2 spectra of HD\,159176, to the three EPIC spectra considered simultaneously and finally to all the {\it XMM} data taken together. The adopted model is {\tt wabs$_\mathrm{ISM}$*wabs$_\mathrm{WIND}$*(mekal + mekal)}. The column density given in the Table refers to the local (i.e.\ circumstellar) absorption. The interstellar column density is frozen at 0.17\,10$^{22}$\,cm$^{-2}$. The error bars represent the 1 $\sigma$ confidence interval.\label{fittable2}}
\begin{center}
\begin{tabular}{ccccccc}
\hline
\hline
	& N$_\mathrm{H}$ & kT$^1$ & Norm$^1$ & kT$^2$ & Norm$^2$ & $\chi^2_\nu$ \\
	& (10$^{22}$ cm$^{-2}$) & (keV) & & (keV) & & d.o.f. \\
\hline
\hline
RGS1 & 0.32 & 0.16 & 1.66 10$^{-2}$ & 0.62 & 3.96 10$^{-3}$ & 1.36 \\
	& $\pm$ 0.11 & $\pm$ 0.01 & $\pm$ 1.70 10$^{-2}$ & $\pm$ 0.04 & $\pm$ 1.81 10$^{-3}$ & 61 \\
\hline
RGS2 & 0.27 & 0.16 & 1.37 10$^{-2}$ & 0.59 & 2.67 10$^{-3}$ & 1.45 \\
	& $\pm$ 0.09 & $\pm$ 0.01 & $\pm$ 0.26 10$^{-2}$ & $\pm$ 0.03 & $\pm$ 1.05 10$^{-3}$ & 71 \\
\hline
EPIC & 0.40 & 0.21 & 2.19 10$^{-2}$ & 0.97 & 1.48 10$^{-3}$ & 2.41 \\
	& $\pm$ 0.14 & $\pm$ 0.02 & $\pm$ 0.23 10$^{-2}$ & $\pm$ 0.01 & $\pm$ 0.05 10$^{-3}$ & 726 \\
\hline
EPIC & 0.40 & 0.21 & 2.22 10$^{-2}$ & 0.96 & 1.54 10$^{-3}$ & 2.28 \\
+ RGS	& $\pm$ 0.01 & $\pm$ 0.02 & $\pm$ 0.22 10$^{-2}$ & $\pm$ 0.01 & $\pm$ 0.05 10$^{-3}$ & 862 \\
\hline
\hline
\end{tabular}
\end{center}
\end{table*}

\section{The RGS spectrum of HD\,159176 \label{sect_rgs}}
\begin{figure}
\begin{center}
\resizebox{8.5cm}{5.0cm}{\includegraphics{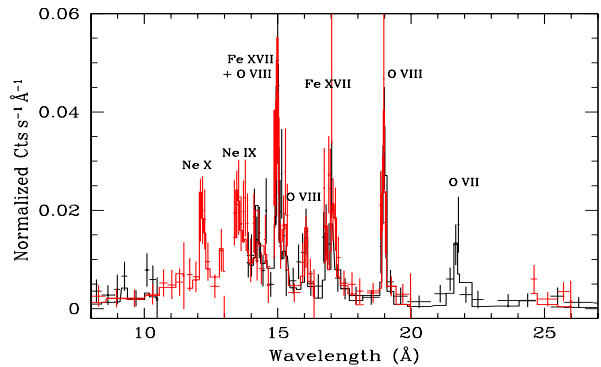}}
\end{center}
\caption{Combined fit of RGS1 and RGS2 spectra between 8 and 27\,\AA\, for a two-temperature {\tt mekal} model. The strongest lines are labelled.\label{rgsspec}}
\end{figure}

Because of the strong soft proton activity and since it turned out to be impossible to merge event lists from separate RGS exposures, we had to restrict our analysis to the results of the pipeline processing of the 16\,ks exposure. We extracted first order spectra of HD\,159176 via the `rgsspectrum' {\sc sas} task. Figure\,\ref{rgsspec} shows RGS1 and RGS2 spectra. The most prominent features in the RGS spectrum are the Ly$\alpha$ lines of \ion{Ne}{x} and \ion{O}{viii}, as well as the He-like triplet of \ion{O}{vii} and some strong \ion{Fe}{xvii} lines. The \ion{Ne}{ix} He-like triplet is also present, but the poor quality of our data as well as the blend with many iron lines from various ionization stages hamper the analysis of this part of the spectrum. The temperatures of maximum emissivity of the lines seen in the RGS spectra of HD\,159176 span a range from $\sim 2$ to $\sim 6\,10^{6}$\,K (i.e.\ kT $\sim$ 0.17 -- 0.52\,keV, see the APED database, Smith \& Brickhouse \cite{SB}).\\

Models were fitted to the RGS spectra. The model including the two-temperature {\tt mekal} component yields the best fits. The combined fit of RGS1 and RGS2 spectra by a two-temperature {\tt mekal} thermal model is also illustrated in Fig.\,\ref{rgsspec}. The best fitting parameter values are given in Table\,\ref{fittable2}. With reasonable $\chi^2_\nu$ values, we obtain temperatures (about 0.16 and 0.60\,keV) consistent with the fit parameters of the EPIC pn spectrum. It is interesting to note that these temperatures correspond to the extreme values of the range inferred hereabove from the maximum emissivities of the lines identified in the spectrum. The interstellar column density appears also comparable within the error bars to the EPIC results for the same two-temperature model.

The agreement between the RGS and EPIC pn parameters suggests that we can attempt a simultaneous fit of the spectra from different instruments. For this purpose, we used the two-temperature {\tt mekal} model. The results are summarized in Table\,\ref{fittable2}. First, we fitted the three EPIC spectra together. Next, we also included the RGS spectra.

We see that, in the first case, best fitting parameter values are very close to the results obtained hereabove for the MOS spectra (which are somewhat different from the best fitting parameters of the pn spectrum). On the other hand, including also the RGS spectra in the fit, we obtain a value of the second temperature which is also in good agreement with that obtained for the MOS spectra fitted individually. Finally, we note that fitting the EPIC + RGS data with a differential emission measure model (i.e.\ the {\tt c6pmekl} model in {\sc xspec}, see Singh et al.\, \cite{singh}) yields a differential emission measure with a very broad maximum spanning from about 0.15 to 1.2\,keV, again in reasonable agreement with the temperature ranges inferred from the RGS line spectrum.\\
The best fitting temperatures of the X-ray spectra of HD\,159176 are rather low. In fact, these temperatures are roughly typical for presumably single O-stars. On the other hand, for an O + O colliding wind binary, one expects a priori to find harder X-ray emission from the wind interaction region. For instance, in the spectrum of HD\,93403, Rauw et al.\ (\cite{hd93403}) found a hot thermal component with kT $\sim$ 0.92 -- 1.25\,keV, which is significantly hotter than the temperature of the harder component in the EPIC and RGS spectra of HD\,159176. However, the smaller orbital separation of HD\,159176 (as compared to HD\,93403) prevents the stellar winds of its components from reaching their terminal velocities before they collide, leading to lower temperatures in the wind interaction region.\\

\begin{figure}
\begin{center}
\resizebox{8.5cm}{6.3cm}{\includegraphics{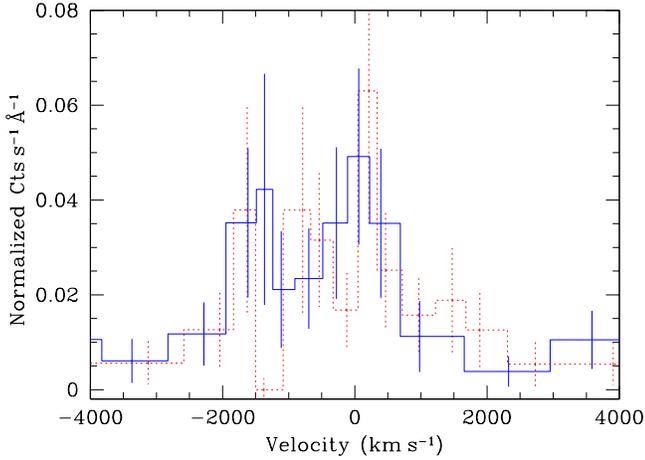}}
\end{center}
\caption{The O\,{\sc viii} Ly\,$\alpha$ line at 18.97\,\AA\ as seen with the RGS1 (dotted line) and RGS2 (solid line) instrument.\label{lyalox}}
\end{figure}

Given the rather poor signal-to-noise ratio of our RGS data, nothing reliable can be said about the morphology of the emission lines. For instance, the strongest line of the spectrum, the O\,{\sc viii} Ly\,$\alpha$ line at 18.97\,\AA, is illustrated in Fig.\,\ref{lyalox}. This line appears to be significantly broadened. In fact, the line's FWHM estimated from Fig.\,\ref{lyalox} is about 2500\,km\,s$^{-1}$, while the FWHMs of the instrumental profiles at this wavelength are about 0.06 and 0.07\,\AA\ (i.e.\ 950 and 1100\,km\,s$^{-1}$) for RGS1 and RGS2 respectively (den Herder et al.\ \cite{RGS}). The line appears therefore significantly larger than what would be expected for a coronal emission model (see Owocki \& Cohen \cite{OC}). We also note that the line profile is neither gaussian nor lorentzian. Theoretical models of line emission from either single O-type stars or colliding wind binaries (e.g.\ Owocki \& Cohen \cite{OC}, Feldmeier et al.\ \cite{feld2}, Henley et al.\ \cite{HSP}) predict rather flat-topped profiles with asymmetries depending on the amount of wind absorption and, in the case of binary systems, on the orbital phase. The detection of an orbital modulation of the X-ray line profiles such as predicted by Henley et al.\ would provide a clear evidence that the lines are formed in the wind collision zone. However, since we have only a single {\it XMM-Newton} observation of HD\,159176 at our disposal, we cannot make any statement about the existence or not of such a modulation.
Unfortunately, our current data set does not have the quality needed to perform a detailed comparison between the observed lines and the predictions from the various models. All we can state is that the O\,{\sc viii} Ly\,$\alpha$ line is slightly blue-shifted by about 300 -- 600\,km\,s$^{-1}$, though the highest peak in the profile is found near the rest wavelength. Within the error bars, these features are in reasonable agreement with some of the line profiles predicted by Owocki \& Cohen (\cite{OC}) for a single O star wind. On the other hand, the agreement with line profiles formed in a wind interaction zone for a wind momentum ratio of $\eta \sim 1$ (see below) and an orbital phase of 0.53 -- 0.66 (corresponding to $\theta \in [98^{\circ},130^{\circ}]$ in the formalism of Henley et al.\ \cite{HSP}) is slightly less good. At these phases, the colliding wind models predict the highest peak to occur with a significant blue-shift (Henley et al.). We caution however that the line profiles presented by Henley et al.\ were computed assuming an adiabatic wind interaction, while the wind interaction zone in HD\,159176 is most likely radiative (see below).

Another noteworthy feature of the RGS spectra of HD\,159176 is the fact that the forbidden ($f$) component of the O\,{\sc vii} He-like triplet is very weak. Inspection of the spectrum reveals no significant excess of counts above the level of the pseudo-continuum at the wavelength of the $f$ line, whereas the intercombination ($i$) and resonance ($r$) lines are clearly detected at a level of about $7 \times 10^{-5}$ and $3 \times 10^{-5}$\,photons\,cm$^{-2}$\,s$^{-1}$ respectively. Unfortunately, the limited quality of our data does not allow us to perform sophisticated fits of the triplet. Nevertheless, we have fitted the triplet by three gaussians plus a constant (to mimic the pseudo-continuum) where we requested that the three gaussians should have the same Doppler shift and the same width. This fit yields an $f/i$ ratio of 0.08 which can probably be considered as a safe upper limit. 
A low $f/i$ ratio is not unexpected. In fact, in the winds of hot early-type stars, the intense UV radiation induces a strong coupling between the upper level of the forbidden line and the upper level of the intercombination line, leading to an enhancement of the $i$ and a decrease of the $f$ component (Porquet et al.\ \cite{Por}). For the O\,{\sc vii} He-like triplet, Porquet et al.\ calculate an $f/i$ ratio of less than $3 \times 10^{-3}$ for a star with a radiation temperature $\geq 30\,000$\,K and for a dilution factor $w \geq 0.1$. In HD\,159176, this mechanism can act on the intrinsic X-ray emission of the individual binary components as well as on the X-ray emission from the colliding wind region. Indeed, assuming that the wind-interaction zone lies about half-way between the two stars, the UV radiation field of each star is diluted by $w = 0.15$ at the location of the colliding wind region. The enhanced plasma density in the colliding wind region could further reduce the $f/i$ ratio (Porquet et al.\ \cite{Por}), though this is likely to be only a second order effect in the case of a hot, close binary such as HD\,159176. \\

Finally, a further comparison of the RGS spectrum (Fig.\,\ref{rgsspec}) with RGS spectra obtained for stars such as 9\,Sgr (Rauw et al.\,\cite{9sgr}) or $\zeta$\,Pup (Kahn et al.\,\cite{Kahn}) shows that the HD\,159176 spectrum does not display the \ion{N}{vii} $\lambda$ 24.8\,\AA\, line. To find out whether this is due to a reduced N abundance or not, we fitted simultaneously RGS1 and RGS2 spectra with variable abundance {\tt mekal} models, allowing the nitrogen abundance to vary (with other abundances frozen at the solar value). The resulting fit suggests a slightly (2.85 $\pm$ 1.26) enhanced N abundance (with respect to solar abundance) while we recover the temperatures for the RGS fits within the error bars. Moreover, the \ion{N}{vii} $\lambda$ 24.8 \AA\, line appears clearly in the theoretical RGS spectrum of HD\,159176 when the absorption columns are set to zero, albeit this line nearly completely disappears when the absorption components are set to the values derived in our fit. Therefore, the absence of the \ion{N}{vii} line in our observed spectrum is due to the strong absorption that photons at this rather low energy suffer when they encounter the circumstellar or interstellar medium. Also, the relative intensities of oxygen lines compared to other elements seem quite higher in the spectrum of HD\,159176. At this stage, it is worth recalling that $\zeta$\,Pup is an evolved object and the line intensities in its RGS spectrum reflect the signature of the CNO process (Kahn et al.\,\cite{Kahn}). On the other hand, we do not expect to see such an effect in the spectrum of a relatively unevolved system such as HD\,159176.

\section{Short-term variability of the X-ray flux}
Hydrodynamic models of colliding wind binaries predict the existence of various kinds of dynamical instabilities that should affect the wind interaction region (Stevens et al.\ \cite{SBP}, Pittard \& Stevens \cite{PS}) and manifest themselves through a rapid variability of the emerging X-ray flux in addition to the much slower orbital phase-locked modulation. Theoretical considerations (Stevens et al.\ \cite{SBP}) indicate that the nature of these instabilities depends on the efficiency of radiative cooling in the shocked region which is a function of the orbital separation and of the wind properties of the components of the binary system. Previous X-ray satellites lacked the required sensitivity to look for the signature of these variations. Owing to the large collecting area of the X-ray mirrors onboard {\it XMM-Newton} and since HD\,159176 is rather X-ray bright, we can use our present data set to seek for the signature of these instabilities.

\begin{figure}[h]
\begin{center}
\resizebox{8.5cm}{4.0cm}{\includegraphics{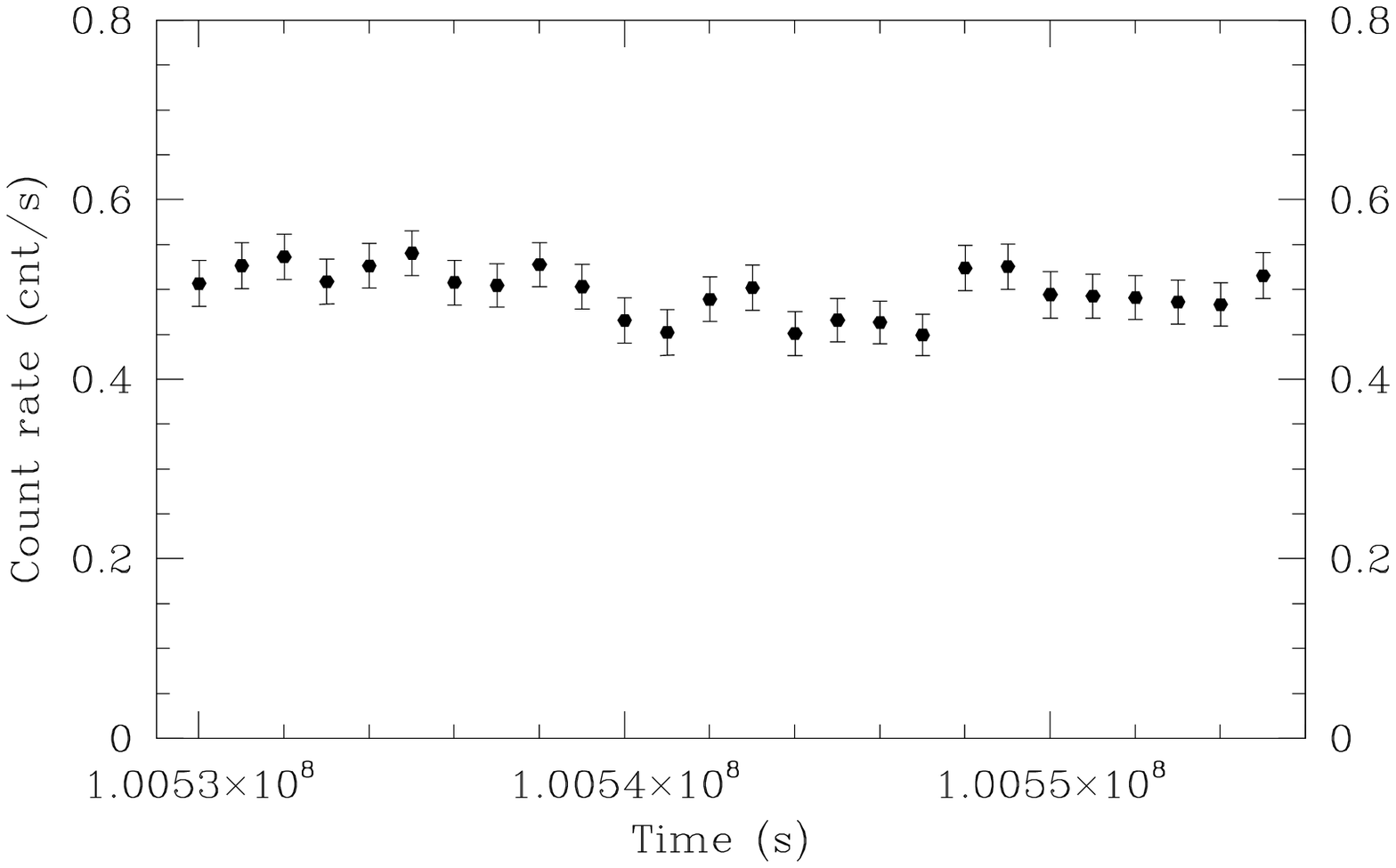}}
\end{center}
\begin{center}
\resizebox{8.5cm}{4.0cm}{\includegraphics{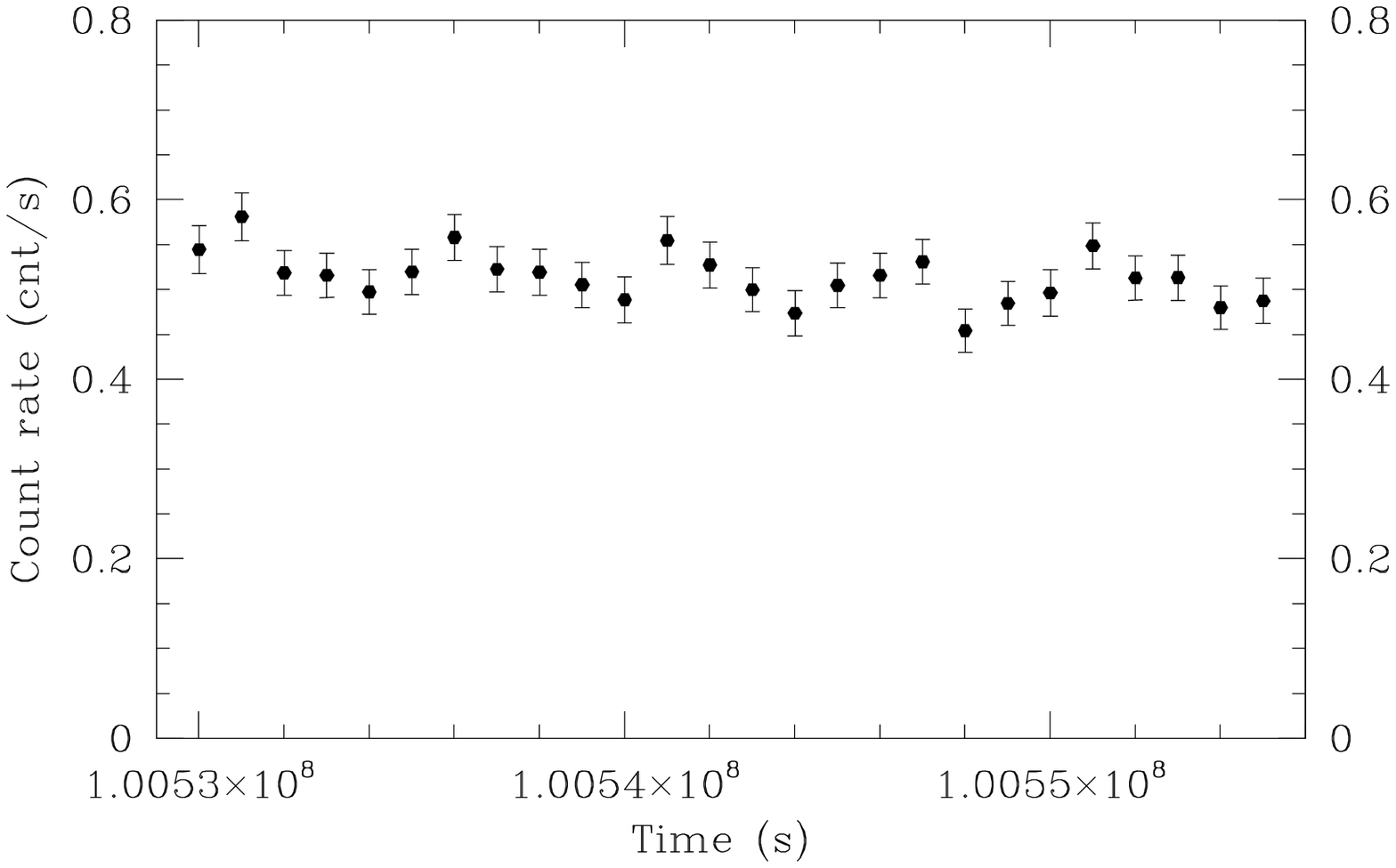}}
\end{center}
\begin{center}
\resizebox{8.5cm}{4.0cm}{\includegraphics{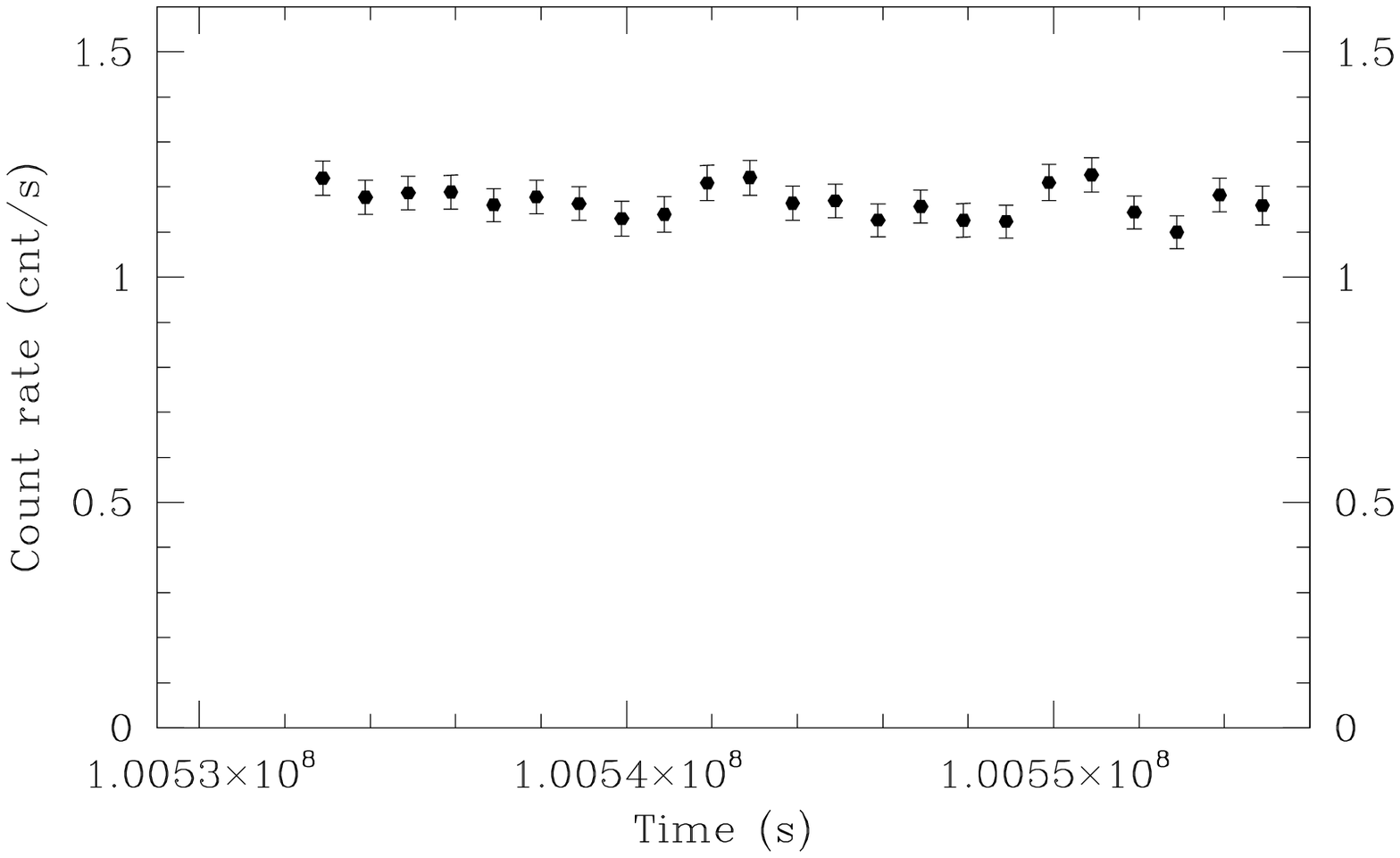}}
\end{center}
\caption{Count rates evaluated in the 0.4-10.0 keV energy band versus time for MOS1 (upper panel), MOS2 (middle panel) and pn (bottom panel). The time bin used is 1000 s for all instruments. The error bars represent the $\pm$ 1\,$\sigma$ standard deviation.\label{variabcr}}
\end{figure}

To investigate the putative short-term variability of HD\,159176, we divided the effective exposure into short time intervals. In the case of the MOS instruments, we used 2.1\,ks time intervals. We obtained 9 and 10 equal time intervals respectively for MOS1 and MOS2. Because the good time intervals (GTIs) of the different instruments are not strictly identical, the time bins of the time series of the two instruments do not match exactly. A spectrum was extracted for each time bin through the same procedure as described in Section\,\ref{sect_epic}. Each spectrum was rebinned, to get at least 4 counts per energy channel, and fitted with a two temperature {\tt mekal} model.\\

Some parameters seem indeed to change with time but no correlation is found between MOS1 and MOS2. In each case, the lower temperature does not vary significantly while the higher temperature component displays larger variations. No systematic trend is found for these variations. The same study was carried out for EPIC-pn data using the two models described in Sect.\,3 for time intervals of 2 ks. In this case, we do not observe any significant variation in the model parameters either.\\

A more model-independent way to check for variability consists in studying the run of count rates for particular time bins, over different energy ranges: see for instance Fig.\,\ref{variabcr} for count rates between 0.4 and 10.0 keV within 1 ks time bins. This figure does not reveal any significant correlation between the different EPIC instruments. The same conclusion applies also for narrower energy ranges selected within this same interval. Moreover, we see that count rates in all energy bands are nearly constant within 2 $\sigma$ for all instruments.

Finally, we extracted light curves for the three EPIC instruments with smaller time bins of 200\,s and 100\,s, within spatial regions corresponding to the source and the background areas selected for spectra extraction (Sect.\,\ref{sect_epic}). Because of standard GTIs, some of the time intervals were shorter than 200\,s (or 100\,s) and we therefore scaled the count rates to the effective exposure time within each bin. The background corrected count rates display some fluctuations but do not reveal any correlation between the various instruments. $\chi^2$ tests performed on these corrected light curves confirm this lack of short time scale variability.

\section{Confrontation with theoretical models}
To complete our study, we can compare the overall properties of the observed spectrum of HD\,159176 with the predictions of colliding wind models.

\subsection{Analytical and hydrodynamical estimates of the X-ray luminosity}

A relevant quantity which should be introduced here is the cooling parameter: $$\chi = \frac{t_\mathrm{cool}}{t_\mathrm{esc}} = \frac{v^4_3\,d_7}{\dot{\rm M}_{-7}}$$ where $t_\mathrm{cool}$, $t_\mathrm{esc}$, $v_3$, $d_7$, and ${\dot{\rm M}_{-7}}$ are respectively the cooling time, the escape time from the shock region near the line of the centers, the pre-shock velocity in units of 1000\,km\,s$^{-1}$, the distance from the stellar center to the shock in units of 10$^{7}$\,km, and the mass loss rate in units of 10$^{-7}$\,M$_\odot$\,yr$^{-1}$ (see Stevens et al.\,\cite{SBP}). Depending on the binary parameters, the gas in a colliding wind region could be either adiabatic ($\chi$\,$\geq$\,1) or radiative ($\chi$\,$<$\,1). While present day hydrodynamical models probably provide a safe estimate of the X-ray emission in adiabatic winds, their use in a highly radiative situation is more problematic. In fact, for values of $\chi$ lower than about 2 or 3, the numerical simulations fail to spatially resolve the cooling length in the wind interaction region. Moreover, the thin-shell instabilities that appear in the simulation of these shocks introduce a mixing of hot and cool material that further complicates the prediction of the X-ray emission. Therefore, in the radiative case, it seems more appropriate to use analytical results to estimate the X-ray luminosity.\\

Turning back to HD\,159176, we can estimate the value of the relevant cooling parameter using the binary parameters given in Table\,\ref{tbl-1a}. In a close binary such as HD\,159176, the stellar winds do not reach v$_\infty$ before they collide. The pre-shock velocity should be in the range 850 -- 2000 km\,s$^{-1}$. The upper boundary of this range is obtained assuming a `standard' wind velocity law with $\beta = 0.6$ and v$_{\infty} = 2850$\,km\,s$^{-1}$; the lower boundary corresponds to the value expected if we account for radiative inhibition effects for equal stars with a separation of about 40\,R$_\odot$ (see Fig.\,4 of Stevens \& Pollock \cite{SP}). These pre-shock velocity values are derived along the line connecting the centers of the stars, where the bulk of the X-ray emission is generated.\\

Therefore, adopting the parameters of Table\,\ref{tbl-1a} and a pre-shock velocity lower than 2000 km\,s$^{-1}$, the resulting value of $\chi$ is typical of radiative winds. The corresponding intrinsic X-ray luminosity can be estimated for each component of the binary through the following relation taken from  Pittard \& Stevens (\cite{PS2}): $$ L_\mathrm{X, intr} \sim 0.5\,\Xi\,{\dot{\rm M}}\,v^2 $$ where $\Xi$ accounts for a geometrical factor (1/6 for equal winds) and a small inefficiency factor (hence we used $\Xi$ = 0.1), and $v$ is the wind speed at the contact surface. In the conditions described hereabove, L$_\mathrm{X,intr}$ is significantly larger than our observed value. For a pre-shock velocity of 2000\,km\,s$^{-1}$, L$_\mathrm{X,intr}$ ($\approx 8\,10^{35}$\,erg\,s$^{-1}$) is more than a factor 400 too high. Note that the intrinsic luminosity needs to be corrected for stellar wind absorption before it can be compared to the observed luminosity. However, as we will see below, this correction hardly exceeds a factor of a few tens and is therefore not sufficient to account for such a huge discrepancy. If we reduce the pre-shock velocity to 850\,km\,s$^{-1}$, the intrinsic luminosity becomes L$_\mathrm{X,intr} \approx$ 10$^{35}$ erg\,s$^{-1}$, which is still much too large.\\

In order for our predicted luminosity to match the observed value of L$_\mathrm{X}$, we need to reduce the mass-loss rates. As a second step, we therefore adopt ${\dot{\rm M}} = 6\,10^{-7}$\,M$_\odot$\,yr$^{-1}$ for both stars as suggested by Howarth \& Prinja (\cite{HP}). With a pre-shock velocity of 2000 km\,s$^{-1}$, $\chi$ amounts to $\sim$\,3.5, allowing us to derive the X-ray flux from a hydrodynamical simulation. For this purpose, we used the code of Pittard \& Stevens (\cite{PS}) which is based on the VH-1 numerical code (see Blondin et al.\,\cite{Blondin}). This code uses a lagrangian piecewise parabolic method to solve the partial differential equations of hydrodynamics followed by a remap onto a fixed grid after each time step (for details see the references in Pittard \& Stevens \cite{PS}). The simulations do not include radiative driving and assume instead constant wind speeds. A theoretical attenuated X-ray spectrum was generated from a 3D radiative transfer calculation on the hydrodynamical results, and we found L$_\mathrm{X,att}$ $\sim$ 2\,10$^{34}$ erg\,s$^{-1}$. Hence it seems that $\dot{\rm M}$ must be reduced below the Howarth \& Prinja (\cite{HP}) value in order to account for the observed L$_\mathrm{X}$.\\ 

In order to explore a wider range of the wind parameter space, we decided to test an alternative approach that allows us to overcome some of the practical limitations encountered hereabove. This is described in the following section.\\

\subsection{Steady-state colliding wind model}
The main problem when modelling the X-ray emission from a close binary system like HD\,159176 is that the wind-wind collision can be highly radiative. Since it is difficult, if not impossible, to resolve the width of such shocks in hydrodynamical calculations, the necessary information to compute the X-ray emission is lost. For this reason, Antokhin et al. (\cite{Ant}, in preparation) have developed a steady-state model able to account for the physics of radiative shocks in a way suitable to lead to a relevant discussion of colliding wind X-ray emission from close binary systems.\\

In this approach, Antokhin et al. (\cite{Ant}) completely neglect instabilities and orbital rotation.  Assuming a strong radiative cooling, both shocks generated in the primary and the secondary winds are thin. The pre-shock wind velocities are specified by a $\beta-$law, and the shape of the contact surface is computed. The wind density and normal velocity are calculated at the exact position of the shock. The 1D hydrodynamic equations are then solved for the case of a radiative post-shock flow. The X-ray spectrum is computed, and the process is repeated along the contact surface to obtain an integrated spectrum. The absorption of the X-ray radiation by the winds and by the shocks is determined using warm material opacity tables obtained with the {\sc cloudy} v.94.00 code (http://thunder.pa.uky.edu/cloudy/). This model neglects any dissipative process and assumes that all the kinetic energy of the winds is radiated, thus giving an upper limit to the actual X-ray luminosity. For further details about this model, we refer the reader to Antokhin et al. (\cite{Ant}).\\ 

\subsubsection{Equal winds}

\begin{figure}
\begin{center}
\resizebox{8.5cm}{9.0cm}{\includegraphics{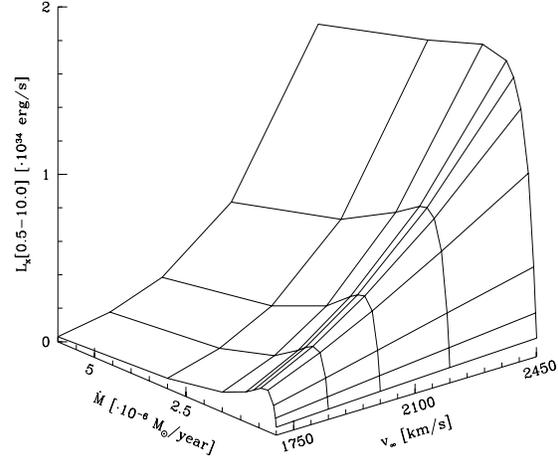}}
\end{center}
\caption{Emerging X-ray luminosity between 0.5 and 10.0 keV as a function of the mass loss rate and the terminal velocity as computed for a distance of 1.5 kpc. The parameter space covers values between 2.5\,10$^{-8}$ and 6.0\,10$^{-6}$ M$_\odot$\,yr$^{-1}$ for the mass loss rate, and between 1700 and 2450 km\,s$^{-1}$ for the terminal velocity.\label{parsp}}
\end{figure}

Following this approach, a grid of models was generated and converted into a table model suitable to be used within {\sc xspec}. As a first step, the assumption of equal winds was adopted. The parameters of the model are the mass loss rate and the terminal velocity. The parameter space covers ${\dot{\it {\rm M}}}$ values from 2.5\,10$^{-8}$ to 6.0\,10$^{-6}$ M$_\odot$\,yr$^{-1}$, and v$_\infty$ values between 1700 and 2450 km\,s$^{-1}$. The luminosities computed by the model were converted into fluxes assuming a distance of 1.5 kpc. The model normalization which scales as the inverse of the distance squared should be unity if all the model assumptions are correct and the distance is indeed equal to 1.5 kpc. Fig.\,\ref{parsp} shows the emerging (i.e. after absorption by circumstellar material) luminosity between 0.5 and 10.0 keV versus the mass loss rate and the terminal velocity. This figure shows clearly that the general trend followed by the luminosity is to decrease as the velocity decreases. For ${\dot{\it {\rm M}}}$ values lower than about 8.5\,10$^{-7}$ M$_\odot$\,yr$^{-1}$, L$_\mathrm{X}$ increases as the mass loss rate increases, but it decreases with increasing mass loss rate for higher ${\dot{\it {\rm M}}}$ values. This turnover in the behaviour of the luminosity versus ${\dot{\it {\rm M}}}$ is due to the increase of the optical depth of the warm wind material as the wind density increases.\\

\begin{figure}
\begin{center}
\resizebox{8.5cm}{6.3cm}{\includegraphics{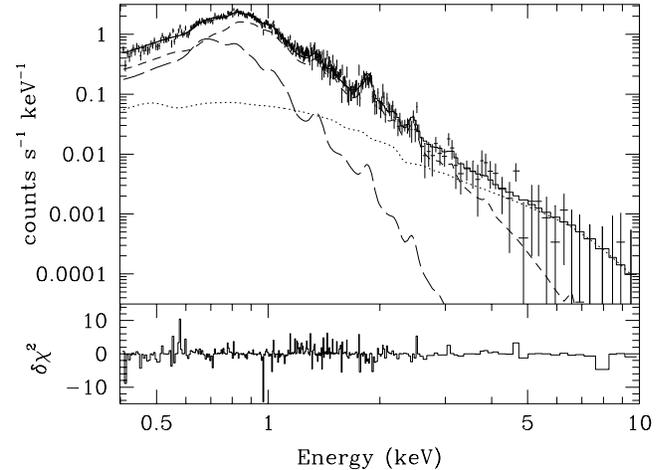}}
\end{center}
\caption{EPIC-pn spectrum of HD\,159176 fitted with a {\tt mekal} + colliding wind + power law model between 0.4 and 10.0\,keV in the case of equal winds. The contributions of the three components are also individually displayed: the long-dashed, short-dashed, and dotted lines are respectively the {\tt mekal}, wind-collision, and power law components. \label{pn_cwb1}}
\end{figure}

To take into account the fact that intrinsic emission of the individual winds is supposed to contribute to the observed spectrum, we used a {\tt mekal} thermal plasma component in addition to the colliding wind component. Interstellar absorption is accounted for by using an absorption column frozen to the value used in Sect.\,\ref{sect_epic}. The best fit of our EPIC-pn spectrum requires an additional power law component to fit the high energy tail already mentioned in the framework of the EPIC-pn spectral analysis in Sect.\,\ref{sect_epic}. The physical motivation of this power law is discussed in Sect.\,\ref{concl}\footnote{We explored higher velocity values to try to fit this high energy tail with the colliding wind model. However, these models yield an increase of the intensity of the Fe K line at about 6.7 keV up to a level that is totally inconsistent with our data.}. Fig.\,\ref{pn_cwb1} shows the fit obtained between 0.4 and 10.0 keV where the three individual components are displayed. The fluxes of the three components in the 0.4--10.0 keV band are 1.5\,10$^{-12}$ erg\,s$^{-1}$, 3.8\,10$^{-12}$ erg\,s$^{-1}$ and 5.7\,10$^{-13}$ erg\,s$^{-1}$ respectively for the {\tt mekal}, the colliding wind and the power law components. This fit corresponds to $\log{\dot{\it {\rm M}}} = -6.67 \pm 0.09$, v$_\infty = 1905 \pm 50$ km\,s$^{-1}$, and a normalization parameter of $0.24 \pm 0.04$. In the best fit model, the {\tt mekal} component has a temperature of 0.29 $\pm$ 0.01 keV, and the power law has a photon index of 2.50 $\pm$ 0.22. While this model provides an acceptable fit to the data -- the reduced $\chi_\nu^2$ is 1.14 for 319 d.o.f. -- we note that the best fit value of the normalization parameter is a factor 4 lower than expected. Because of the membership of HD\,159176 to the open cluster NGC\,6383, it seems quite unlikely that the distance could be off by as much as a factor 2. Moreover, in order to explain the low normalization parameter, the actual distance of HD\,159176 would have to be larger by a factor 2 whereas an estimate of the distance on the basis of the expected absolute magnitude of an O7V + O7V binary rather suggests that the distance of HD\,159176 might be less than 1.5 kpc\footnote{With m$_\mathrm{V}$=5.64 (FitzGerald et al.\,\cite{Fitz}) corrected for the reddening (A$_\mathrm{V}$=1.12) and M$_\mathrm{V}$=--4.8 (Conti \& Alschuler \cite{CA}) for a typical O7V star, leading to M$_\mathrm{V}$=--5.55 for an O7V + O7V system, we infer a distance of about 1.0 kpc only. With M$_\mathrm{V}$=--5.2 (Schmidt-Kaler \cite{SK}), the calculated distance is about 1.2 kpc.}. If the normalization value is frozen to one, the model clearly overestimates the data mostly between 1 and 5 keV, leading to an unacceptable $\chi^2$. Therefore, the low normalization value probably means that the model predicts, for the derived $\log{\dot{\it {\rm M}}}$ and v$_\infty$, a luminosity which is too large by about a factor of 4. A similar result is obtained for a simultaneous fit of MOS1 and MOS2 data with the same model. The only difference is a somewhat higher terminal velocity (v$_\infty = 2190 \pm 50$ km\,s$^{-1}$), compatible with the higher temperature obtained for {\tt mekal} models described in Sect.\,\ref{sect_epic}, and a stronger disagreement of the normalization parameter which is a factor about 8 too low.\\

The results of this subsection can be summarized as follows. Exploring a domain of the parameter space that includes the full range of reasonable values of ${\dot{\it {\rm M}}}$ and v$_\infty$ for stars of spectral type O7 $\pm$ one subclass, we have found a model where the interplay between emission and circumstellar absorption allows to reproduce the observed spectral shape. However, this model as well as all other combinations of ${\dot{\it {\rm M}}}$ and v$_\infty$ in this domain of the parameter space fail to reproduce the observed luminosity. The most likely explanation for this failure is that one or several of the model assumptions break down in the case of HD\,159176. One possibility that we briefly explore in the following section could be that the stars have winds of unequal strengths.\\

\subsubsection{Unequal winds \label{uneq}}
Although Pachoulakis (\cite{Pach}) inferred only slightly different wind momenta (see Table\,\ref{tbl-1a}), we note that the stellar parameters (T$_\mathrm{eff}$, L$_\mathrm{bol}$) derived by Pachoulakis yield very different values of ${\dot{\it {\rm M}}}$ (9\,10$^{-7}$ and 1\,10$^{-7}$ M$_\odot$\,yr$^{-1}$) when passed through the relation of Vink et al. (\cite{Vink}). Such large differences in ${\dot{\it {\rm M}}}$, and hence wind strength, are however at odds with the essentially equal spectral types inferred from optical spectroscopy. Nevertheless, we attempted to establish whether an unequal wind CWB model could reconcile the predicted and observed luminosities. We thus developed a grid of models depending on four parameters: the terminal velocity of the primary (v$_{\infty,1}$), the mass loss rate of the primary (${\dot{\it {\rm M}}}_1$), the terminal velocity of the secondary (v$_{\infty,2}$), and finally the wind momentum ratio defined as $$ \eta = \frac{{\dot{\it {\rm M}}}_1\,v_{\infty,1}}{{\dot{\it {\rm M}}}_2\,v_{\infty,2}} .$$ This latter parameter is used to control the ratio of the wind strengths, to avoid any collision of the primary wind with the photosphere of the secondary. An additional normalization parameter as described for the equal wind case is also used.\\
The velocity parameters are allowed to vary in a range between 1700 and 2500 km\,s$^{-1}$, and the allowed values for ${\dot{\it {\rm M}}}_1$ are taken between 3.0\,10$^{-8}$ M$_\odot$\,yr$^{-1}$ and 6.0\,10$^{-6}$ M$_\odot$\,yr$^{-1}$. The parameter $\eta$ can vary between 1.0 and 1.44. Beyond this upper boundary the stronger wind of the primary crashes onto the secondary surface.\\
The result of the fit seems only weakly sensitive to the value of $\eta$ (note that the range covered is quite narrow). If $\eta$ is frozen to unity and velocities constrained to be equal, the same fitting result as described for the equal wind case is obtained. However, we were not able, with the current model, to obtain a satisfactory fit with a norm parameter close to unity. This result illustrates again the impossibility to reconcile the observed spectrum with the model at its current stage of development, at least within the parameter space we explored.\\

\section{Summary and conclusions \label{concl}}
The main results from our study of the {\it XMM-Newton} data of HD\,159176 are the following:
\begin{enumerate}
\item[-] The study of EPIC and RGS data reveals a soft spectrum which is consistently fitted with a two temperature thermal model. The hottest component is at about 0.6 keV for EPIC-pn and RGS, and about 1.0 keV for EPIC-MOS data. The X-ray luminosity is rather low, with a value of L$_\mathrm{X}$/L$_\mathrm{bol}$ showing a moderate excess ($\sim$ 7) compared to what is expected for isolated stars with the same bolometric luminosity but without any wind-wind interaction.\\
\item[-] Our analysis of RGS data reveals that lines are significantly broadened. For instance, the \ion{O}{viii} Ly $\alpha$ line at 19.0 \AA\, has a full width at half maximum of about 2500 km\,s$^{-1}$, in agreement with X-ray lines originating either from shocks distributed throughout the wind or from a colliding wind zone.\\
\item[-] The study of EPIC light curves failed to reveal any significant variability on time scales of 100 -- 2000\,s. This indicates that, at least in a system like HD\,159176, the hydrodynamic instabilities that might exist in the region of the shocked winds are not able to produce a clear variability of the X-ray emission.\\
\item[-] The EPIC spectrum reveals a high energy tail which can not be fitted by thermal models ({\tt mekal} or colliding wind). This hard X-ray emission component was fitted with a power law with a photon index of about 2.5. Its presence could reflect a non-thermal process such as inverse Compton scattering (Pollock \cite{Pol}, Chen \& White \cite{CW}). The photon index is not too far from the 1.5 value expected for an X-ray spectrum arising from a population of relativistic electrons generated through an acceleration mechanism involving strong shocks. The possibility of a non-thermal X-ray component was already suggested for instance for a system like WR\,110 by Skinner et al. (\cite{Sk}). So far, the most prominent indication of relativistic electrons in stellar winds of early-type stars have been found in the radio domain where a significant fraction of the stars were found to display a non-thermal, probably synchrotron, emission (e.g. Bieging et al.\,\cite{BAC}). In the case of HD\,159176, radio observations failed to reveal such a non-thermal component, providing only an upper limit on the radio flux at 6 cm (Bieging et al.\,\cite{BAC}). However, this does not rule out the possibility that relativistic electrons could be accelerated at the wind collision shock. In fact, the acceleration site would be buried so deeply within the radio photosphere that no synchrotron emission could escape and we would therefore observe HD\,159176 as a thermal radio emitter. \\
\item[-] Besides the non-thermal tail, the observed spectral shape can be consistently reproduced using a steady-state colliding wind model with a mass loss rate value of about 1.7 -- 2.6\,10$^{-7}$ M$_\odot$\,yr$^{-1}$ and a terminal velocity ranging between 1850 and 1950 km\,s$^{-1}$ (or between 2140 and 2240 km\,s$^{-1}$ for EPIC-MOS data). However, this model is unable to predict X-ray luminosities compatible with the observed spectrum of HD\,159176. Theoretical values are systematically higher than the observed X-ray luminosities. This disagreement between theory and observation is discussed hereafter.\\
\end{enumerate}

The disagreement between our {\it XMM-Newton} observation and the theoretical predictions could possibly be explained by several factors. First, let us recall that the kinetic power of the collision should be considered as an upper limit on the X-ray luminosity even in a highly radiative system. For instance, some of the collision energy might be taken away by the shocked gas which ends up with negligible velocity and pressure near the line of centers. In order not to pile up at the `stagnation point', this gas must be advected from the system, and the work needed to lift it out of the gravitational potential of the system could take away energy at the expense of X-ray emission. Second, higher values of the parameter $\eta$ (see the unequal wind case, Sect.\,\ref{uneq}) should be considered. Although the current version of the model is unable to deal with the case where the primary wind crashes onto the secondary photosphere, this scenario should be envisaged. Let us emphasize however that the optical spectrum of HD\,159176 does not provide support for very large values of $\eta$. Third, diffusive mixing between hot and cool material is likely to exist due to the instability of the shock front. As a consequence, the material tends to emit a much softer spectrum (i.e. EUV, or even UV), at the expense of X-rays. Unfortunately, current simulations lack the needed resolution to accurately deal with this mixing. Note that thermal conduction (Myasnikov \& Zhekov \cite{MZ}) is also expected to produce a softer spectrum. Finally, further developments of the current model are needed to address this issue. An improvement of the current steady-state model would be for instance to consider the effect of mechanisms able to lower the predicted luminosities like sudden radiative braking (Gayley et al.\,\cite{GOC}), which can potentially occur every time unequal winds interact in a close binary system. In addition, orbital effects should also be included to study such systems.\\

HD\,159176 is the first short period colliding O + O binary studied with the high sensitivity of the instruments on board the {\it XMM-Newton} satellite. The major point of this system is that the shocks associated with the wind collision are radiative, making them very difficult to simulate with current hydrodynamic models. For the first time, an alternative model has been used to address this case, following a steady-state geometrical approach leading to promising results. To achieve a better understanding of the special case of radiative colliding wind shocks, other close binary systems should be observed in conjunction with further developments of the new theoretical approach followed in this study.

\acknowledgement{Our thanks go to Mathias Ehle ({\it XMM}-SOC) for his help in processing the EPIC data and to Alain Detal (Li\`ege) for his help in installing the {\sc sas}. We wish to thank Andy Pollock (ESA) for discussion. The Li\`ege team acknowledges support from the Fonds National de la Recherche Scientifique (Belgium) and through the PRODEX XMM-OM and Integral Projects. This research is also supported in part by contracts P4/05 and P5/36 ``P\^ole d'Attraction Interuniversitaire'' (SSTC-Belgium). JMP gratefully acknowledges funding from PPARC for a PDRA position. IIA acknowledges support from the Russian Foundation for Basic Research (grant 02-02-17524). This research has made use of the SIMBAD database, operated at CDS, Strasbourg, France and of the NASA's Astrophysics Data System Abstract Service.}

\end{document}